\documentclass[conference]{IEEEtran}
\IEEEoverridecommandlockouts
\usepackage{cite}
\usepackage{amsmath,amssymb,amsfonts}
\usepackage{algorithmic}
\usepackage{algorithm}
\usepackage{graphicx}
\usepackage{textcomp}
\usepackage{xcolor}
\usepackage{subcaption}
\def\BibTeX{{\rm B\kern-.05em{\sc i\kern-.025em b}\kern-.08em
    T\kern-.1667em\lower.7ex\hbox{E}\kern-.125emX}}
\begin{document}

\title{RAG-Empowered LLM-Driven Dynamic Radio Resource Management in Open 6G RAN
}

\author{\IEEEauthorblockN{ Onur Salan\IEEEauthorrefmark{1}\IEEEauthorrefmark{2}, Burak Çırağ\IEEEauthorrefmark{2}, Onur Sever\IEEEauthorrefmark{1}, İbrahim Hökelek\IEEEauthorrefmark{1}, Ali Görçin\IEEEauthorrefmark{1}\IEEEauthorrefmark{2},
Hakan Ali Çırpan\IEEEauthorrefmark{2}}

\IEEEauthorblockA{\IEEEauthorrefmark{1}  {Communications and Signal Processing Research (HİSAR) Lab., T{\"{U}}B{\.{I}}TAK B{\.{I}}LGEM, Kocaeli, Turkey}}

\IEEEauthorblockA{\IEEEauthorrefmark{2} Department of Electronics and Communication Engineering, Istanbul Technical University, {\.{I}}stanbul, Turkey} 

}

\maketitle

\begin{abstract}
Implications of the advancements in the area of  artificial intelligence to the wireless communications is extremely significant, especially in terms of resource management. In this paper, a Retrieval-Augmented Generation (RAG)-empowered Large Language Model (ReLLM)-driven dynamic radio resource management framework for Open Radio Access Network (O-RAN) inspired 6G networks is proposed. The introduced methodology leverages the ReLLM framework to interpret both historical and real-time network data, enabling adaptive control of network slices. The ReLLM is founded on two specialized agents, one is responsible for proactively detecting service level agreement (SLA) violations by continuously monitoring and estimating slice-specific performance metrics, and the other one is responsible for dynamically reallocating physical resource blocks when the SLA violation probability exceeds a pre-defined threshold. The primary objective of this dual-agent design is to minimize unnecessary LLM inference calls while satisfying the SLA requirements of the slices, thereby improving computational and energy efficiency. The proposed ReLLM framework is implemented and validated on an end-to-end O-RAN testbed built upon open-source OpenAirInterface emulators. The experimental results demonstrate that the LLM approach with its reduced token consumption feature maintains a near-zero drop ratio for the low-priority slice while simultaneously satisfying acceptable latency performance for the high-priority slice. The ReLLM-driven design improves reliability and SLA compliance, confirming its practicality for real-world O-RAN testbeds and its potential applicability to future 6G networks.
\end{abstract}


\section{Introduction}

As 6G research and development efforts gain momentum worldwide, there is a growing consensus that 6G will adopt an AI-native design paradigm. To this end, the Radio Access Network (RAN) Intelligent Controller (RIC) has emerged as a key enabler of the Open RAN (O-RAN) architecture for AI-based automation of radio access network functions \cite{ric}. O-RAN leverages both near-real-time and non-real-time RIC components to offer a flexible and software-driven control framework that supports intelligent resource management and dynamic policy optimization. Existing AI-based solutions for RAN control mostly rely on reinforcement learning (RL) approaches, which require extensive offline training to generalize under highly dynamic wireless environments \cite{sever2025,hierar}. For example, a deep reinforcement learning (DRL)-based xApp is  developed to autonomously manage radio resources under SLA guarantees \cite{sever2025}. In another study \cite{hierar}, a multi-agent reinforcement learning (RL) approach is proposed for RAN slicing, where limited radio resources are allocated toward satisfying the SLAs of multiple slices. These models often lack interpretability and adaptability when network conditions or service requirements change. These limitations open new opportunities for integrating Large Language Model (LLM)-driven intelligence into O-RAN architectures to enable explainable, flexible, and service level agreement (SLA)-aware resource management in next-generation wireless networks.

A comprehensive survey has been conducted in the literature that highlights the capabilities of LLMs in network automation, decision-making enhancement, and simplifying the management of ubiquitous connectivity \cite{llmtelecom}. LLMs can be applied to diverse telecommunication scenarios, such as generating network configurations, classifying traffic patterns and network faults, optimizing resource allocation strategies, and predicting future traffic loads \cite{llmenabledtelecom}. Furthermore, it is demonstrated in \cite{telecomgpt} that fine-tuned LLMs outperforms general-purpose models in telecommunication applications. The authors in \cite{oransight} introduce a model of domain-specific RAG-empowered LLM framework, which improves accuracy and energy efficiency for O-RAN-related tasks. Another recent work \cite{oranguide} proposes RAG-empowered  multi-agent framework for intelligent control in O-RAN environments. Similarly, \cite{llmxapp} presents an LLM-driven xApp for adaptive radio resource management, employing intelligent prompting to dynamically optimize resource allocation among network slices with diverse QoS requirements.

\begin{figure*}[t]
\includegraphics[width=1\textwidth]{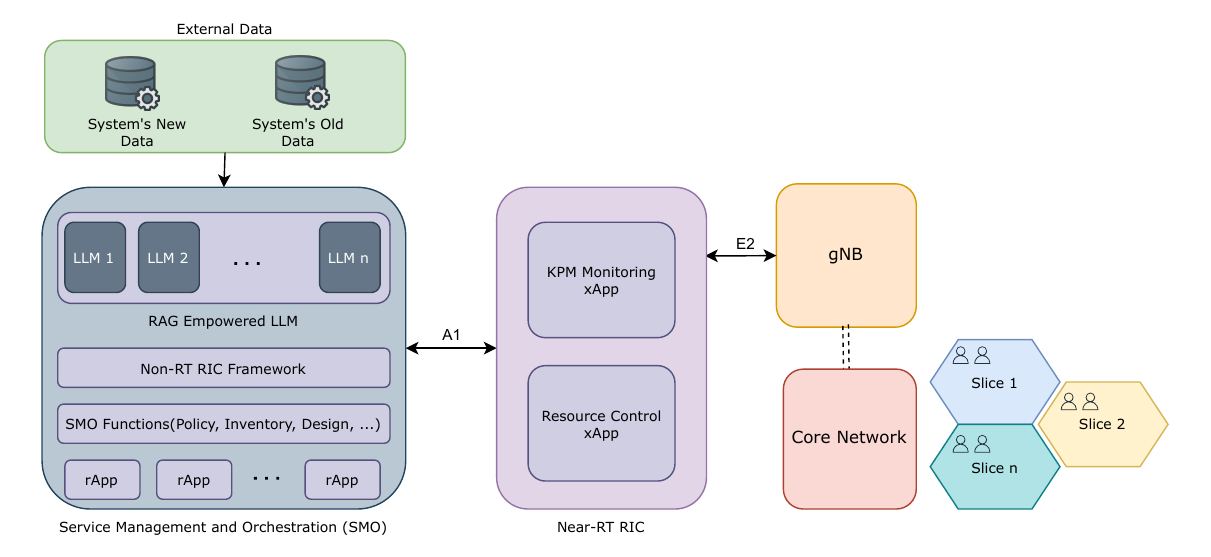}
\centering
\caption{The overview of the proposed O-RAN-based ReLLM-driven dynamic 6G radio resource management framework.}
\label{system_model}
\end{figure*}

While these pivotal studies establish a solid conceptual foundation, the practical realization of LLMs for radio resource management in O-RAN-based networks remains largely unexplored, which serves as the main motivation for the current work. This paper presents a RAG-empowered Large Language Model (ReLLM)-driven approach for dynamic radio resource allocation within O-RAN inspired next generation networks, leveraging both historical and real-time data to adaptively manage network slice resources. The proposed ReLLM system architecture includes two LLM agents; one for proactively detecting SLA violations and the other one for dynamically reallocating physical resource blocks when the SLA violation risk exceeds an administratively defined threshold. The SLA violation detection agent identifies potential SLA violation risks and triggers the resource allocator agent only when necessary, significantly reducing token consumption and inference frequency. A RAG-based architecture stores previous resource allocation decisions along with the corresponding key network performance results. By retrieving relevant historical information through the RAG mechanism, the LLM can make more informed and context-aware decisions without relying on long in-context histories, resulting in improved efficiency and reasoning accuracy. The proposed framework is integrated and validated on an O-RAN inspired end-to-end testbed using open source OpenAirInterface (OAI) emulators, demonstrating its feasibility and effectiveness for practical deployments.

\section{System Model and Problem Formulation}

\subsection{System Model}
Figure \ref{system_model} depicts the system model of an O-RAN inspired network consisting of a core network (CN), a gNB, multiple user equipments (UEs), and an RIC platform. The system includes multiple slices, where each slice serves multiple UEs. The gNB is responsible for managing the finite set of total resource blocks. The CN provides end-to-end connectivity and differentiated services across slices, ensuring that each slice meets its QoS guarantees while maintaining scalability, service isolation, and compliance with SLAs. The RIC is an important enabler in this O-RAN-based architecture, where the AI/ML based management of limited radio resources can be performed using third party applications such as xApps for near-RT RIC and rApps for non-RT RIC. The near-RT RIC hosts two xApps that perform specialized functions such as key performance measurement (KPM) monitoring and slice-aware resource control. The performance metrics are collected from the gNB through the E2 interface and forwarded to the non-RT RIC through the A1 interface. In the proposed model, the service management and orchestration (SMO) framework serves as an upper-level management entity responsible for the orchestration of network and radio resources. The SMO is enhanced with access to external data sources and the proposed ReLLM agents, thereby extending its intelligence and decision-making capabilities. Two specialized LLM agents, namely SLA violation and resource allocation, run in the non-RT RIC platform as rApps.

\subsection{Problem Formulation}
Suppose that there are $N_S$ slices in the network, denoted by $S = \{1,2, \dots, N_S\}$, where each slice serves multiple UEs. Let  $U_k = \{1, 2, \dots, N_{U_k}\}$ denotes the users of the slice $k$, while $N_{RB}$ represents the finite set of total resource blocks. The channel capacity of the user $u \in U_{k}$ at the slice $k$ is represented as:
\begin{equation}
    C_{u,k} = B\sum_{j=1}^{\textit{N}_{RB, u}} \log_2(1+SINR_{u,j})
\end{equation}
where $B$, $\textit{N}_{RB, u}$, and $SINR_{u,j}$ are the bandwidth of each resource block (RB), the number of RBs allocated to the user $u$, and the SINR for the $j^{th}$ RB of the user $u$, respectively. $R_{u,k}$, which represents the throughput of the user $u \in U_{k}$ over the time interval $n$ for the slice $k$, can be calculated as:
\begin{equation}
    R_{u,k}[n] = T_s \times C_{u,k}
\end{equation}
where $T_S$ represents the interval duration. The total throughput of the slice $k$ over the time duration $n$ can be expressed as:
\begin{equation}
    R_k[n] = \sum_{\forall u \in U_{k}}R_{u,k}[n]
\end{equation}

\begin{figure*}[t!]
\includegraphics[width=1\textwidth]{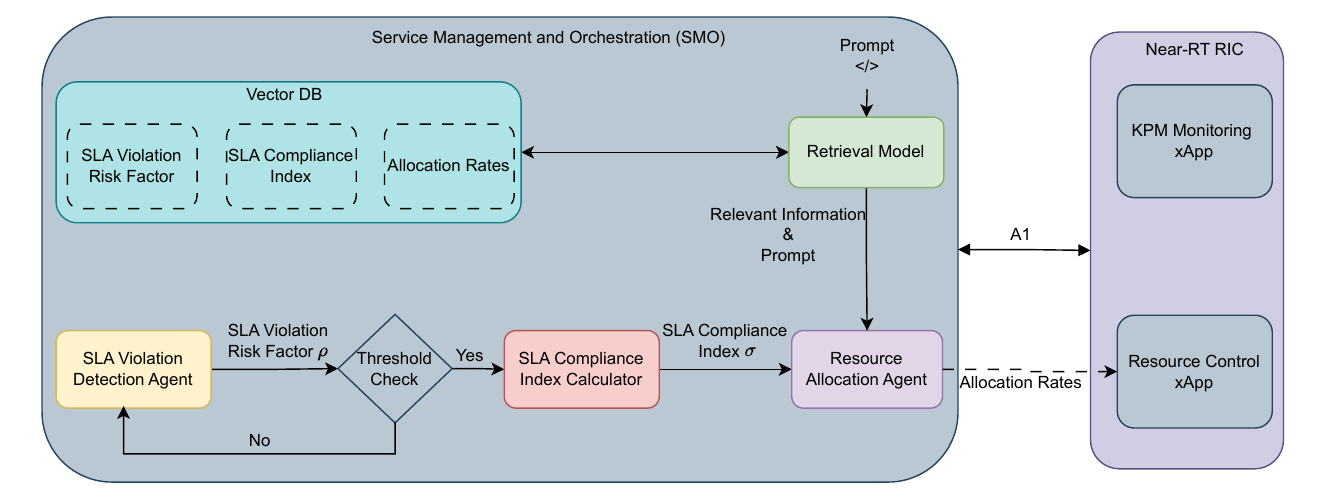}
\centering
\caption{The workflow of the proposed ReLLM-driven framework.}
\label{system_model2}
\end{figure*}

The system model considers the co-existence of latency-constrained  slices, denoted by $\textit{S}_L = \{1, 2, \dots, N_{S_L}\}$, and throughput-constrained slices, denoted by $\textit{S}_T = \{1, 2, \dots, N_{S_T}\}$, where $\textit{S} = S_L \cup S_T$. Each slice serves multiple users, where $U_{L,k} = \{1, 2, \dots, N_{U_{L,k}}\}$ denotes the users associated with the slice $k$ in $\textit{S}_{L}$ with the latency constraint of $\mathcal{L}_{k}$ while $U_{T,k} = \{1, 2, \dots, N_{U_{T,k}}\}$ denotes the users associated with the slice $k$ in $\textit{S}_{T}$ with the throughput constraint of $\mathcal{R}_{k}$. The resource allocation problem aims to maximize the total throughput of the slices in $S_T$ during the time interval of $n$ (i.e., $R_{T}[n]$), while ensuring that the delay requirements of the slices in $S_L$ are satisfied. The optimization problem can be formulated as:

\begin{subequations}
    \begin{equation}
        \textit{maximize } R_{T}[n] = \sum_{\forall k \in S_T }\sum_{\forall u \in U_{T,k}}R_{u,k}[n]
    \end{equation}
    \begin{equation}
        \textit{subject to L}_k[n] = \max_{i \in U_{L,k}}(L_i[n]) < \mathcal{L}_{k}, \hspace{5 mm} \forall k \in S_L
    \end{equation}
        \begin{equation}
        R_k [n]= \min_{i \in U_{T,k}}(R_i[n]) > \mathcal{R}_{k}, \hspace{5 mm} \forall k \in S_T
    \end{equation}
    \begin{equation}
        R_{L}[n] = \sum_{\forall k \in S_L}\sum_{u \in U_{L,k}}R_{u,k}[n]
    \end{equation}
\end{subequations}
where $R_{L}[n]$ denotes the total throughput of the slices in $S_{L}$ within the time interval $n$. Note that this problem is NP-hard and there is no polynomial time algorithm to solve it.  

\subsection{ReLLM Radio Resource Management Framework}
For the sake of simplicity, we assume that there are one latency-constrained slice ($S_1$) and one throughput-constrained slice ($S_2$) in the network. Our approach leverages multi-agent LLM as an intelligent decision-support component that enhances network automation and adaptability in O-RAN system. When combined with RAG, the LLM gains access to both historical and real-time external data, enabling context-aware decision-making for dynamic scenarios such as traffic fluctuations, slice reconfiguration, and SLA compliance. 

The ReLLM module is built upon two specialized agents, including an SLA violation detection agent and a resource allocation agent. The SLA violation detection agent is responsible for detecting potential SLA violations under varying traffic conditions, while the SLA violation risk factor for the slice $k$, denoted by $\rho_k[n]$ over the time interval $n$, is modeled using a sigmoid function as
\begin{equation}
    \rho_k[n]=\frac{1}{1+e^{-a_k(\epsilon_k[n]-b_k)}},
\end{equation}
 where $a_k$ and $b_k$ are slice-specific shaping parameters, and $\epsilon_k$ represents the SLA violation level defined as
\begin{equation}
    \epsilon_k[n] =
    \begin{cases}
        \dfrac{L_{meas}[n] - \mathcal{L}_k}{\mathcal{L}_k}, 
        & k = 1, \\[1.2em]
        \dfrac{R_{meas}[n] - \mathcal{R}_k}{\mathcal{R}_k},
        & k = 2,
    \end{cases}
\end{equation}
where $L_{meas}[n]$ and $R_{meas}[n]$ represent the average latency of $S_1$ and the average throughput of $S_2$ , respectively, computed as the mean of the experimental measurements obtained from the testbed over a given time interval $n$.
The identified SLA violation risk factors are aggregated into a system-wide SLA compliance index, denoted by ($\sigma[n]$), which quantifies the overall network performance in terms of SLA satisfaction over a duration of $n$. The index is formulated as a negative weighted cost function as follows:
\begin{equation}
    \sigma[n] =-\sum_{\forall k \in S}^{}w_k{\rho_k[n]}^2,
\end{equation}
where $w_k$ denotes the priority weight of the slice $k$. The negative sign ensures that the index decreases as the overall SLA violations increase, implying that a higher (less negative) $\sigma[n]$ indicates better SLA compliance and improved system performance. The system reaches its optimal condition when $\sigma[n] = 0$, corresponding to zero SLA violations across all slices. The resource allocation agent dynamically determines the proportion of resources assigned to each slice. Through this hierarchical design, the ReLLM module combines predictive SLA violation analysis with adaptive resource management, thereby enhancing the reliability and efficiency of the O-RAN system. 

\subsection{Workflow of the ReLLM Framework}
    
The dynamic resource allocation process, illustrated in Fig. \ref{system_model2}, begins by initializing the system with an initial resource allocation ratios and a predefined SLA violation threshold $\theta$. KPM Monitoring xApp continuously collects real-time performance metrics such as latency and throughput which are transferred from Near-RT RIC to ReLLM using A1 interface. These metrics are stored in the database and periodically retrieved by the SLA violation detection agent. During a time interval of $n$, the detection agent evaluates the SLA violation risk factors $\rho_k[n]$ for all slices based on the latest KPM data. If all $\rho_k[n]$ values remain below the threshold $\theta$, the system is considered stable, and no reallocation is required. However, if at least one of the $\rho_k[n]$ values exceeds $\theta$ across the time duration $n$, this indicates a potential SLA violation for one or more slices. In such cases, the framework proceeds with dynamic resource reallocation. First, the SLA compliance index ($\sigma[n]$) is computed to determine the current system performance during the time interval. Then, the framework retrieves from the database the examples with the best historical performance and the closest traffic arrival rates, according to the Euclidean distance \cite{faiss}, to provide contextual knowledge. Using the obtained parameters, including $L_{meas}[n]$, $\rho_k[n]$, $\sigma[n]$, and the retrieved examples, a meta-prompt is constructed and provided to the ReLLM resource allocation agent. The LLM processes this prompt and generates a new resource allocation rates, optimizing the distribution of RBs among network slices.
The resulting allocation ratios are then transmitted to the Resource Control xApp, which interacts with gNB through E2 interface where the decisions are applied in real time. The system subsequently observes the new network state, stores updated performance metrics in the database, and repeats the process after a waiting period of $\tau$ seconds.

\section{Implementation and Performance Evaluation}
    \subsection{Experimental Setup and Cases} 
    
    The proposed framework is experimentally validated on an end-to-end 5G O-RAN compliant OAI testbed enhanced with a RAG-empowered LLM, which employs LLaMa 4 Maverick model as the language model, with inference executed on the GroqCloud platform. The FlexRIC platform serves as the RIC implementation, enabling the execution of xApps and rApps for monitoring and resource control, as well as supporting the A1 and E2 interfaces for coordination between the Near-RT and Non-RT RIC \cite{schmidt2021flexric}. 
    The OAI testbed is deployed with two slices consisting of a latency-constrained slice ($S_1$) for delay-sensitive applications, and a throughput-constrained slice ($S_2$) for best-effort traffic.
    At the initial stage, one UE was assigned to each slice, with equal RBs allocation maintained between them. In the proposed system, the SLA is defined to ensure that the latency of the $S_1$ remains below $10$ ms while utilizing the minimum required number of RBs, thereby allowing the remaining radio resources to be dynamically reallocated to $S_2$ to maximize its overall performance. 
    
    \begin{figure}
        \centering
        \begin{subfigure}{\columnwidth} 
            \includegraphics[width=\textwidth, ]{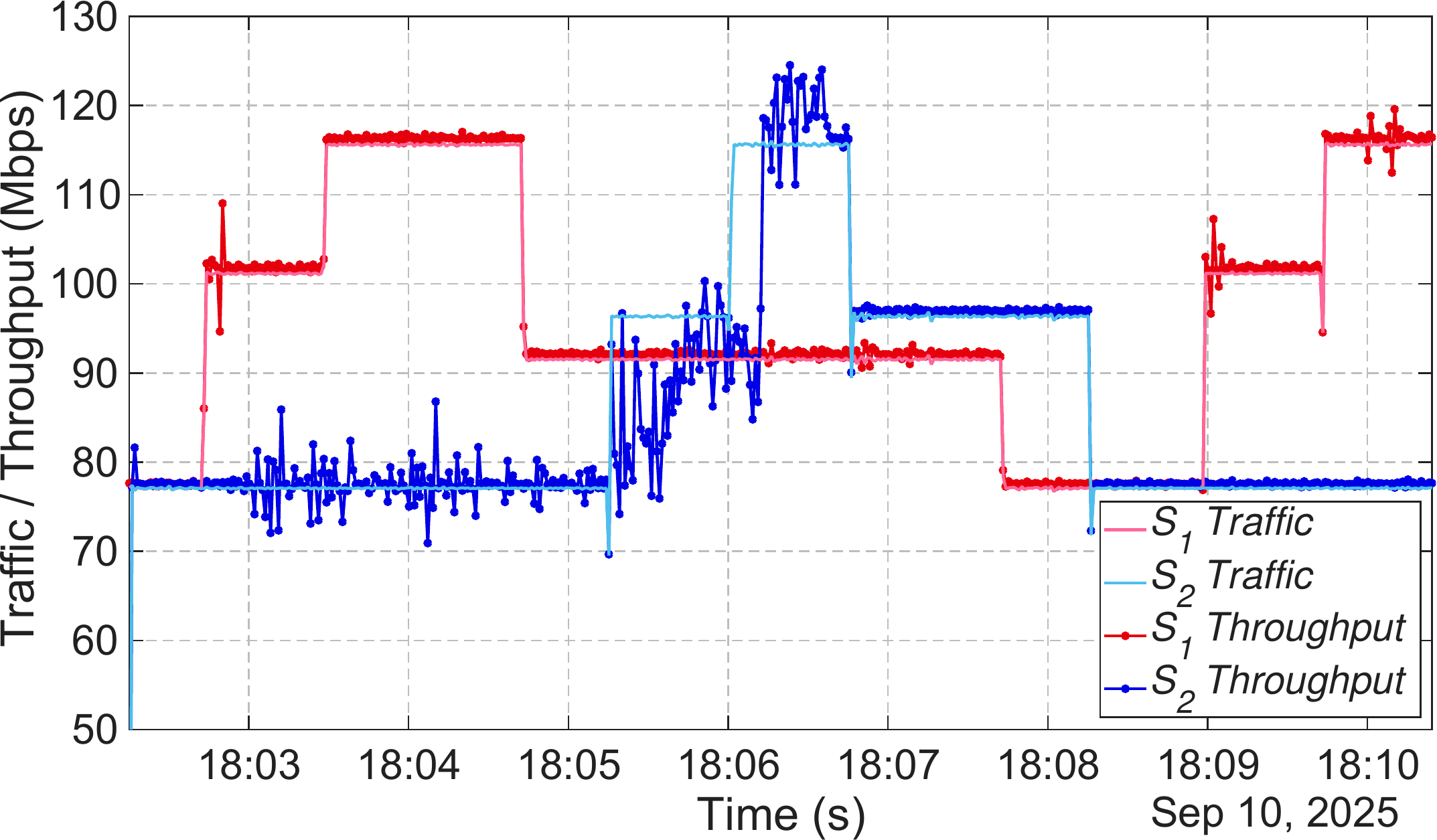}
        \end{subfigure}
        \begin{subfigure}{\columnwidth}
            \includegraphics[width=\textwidth, ]{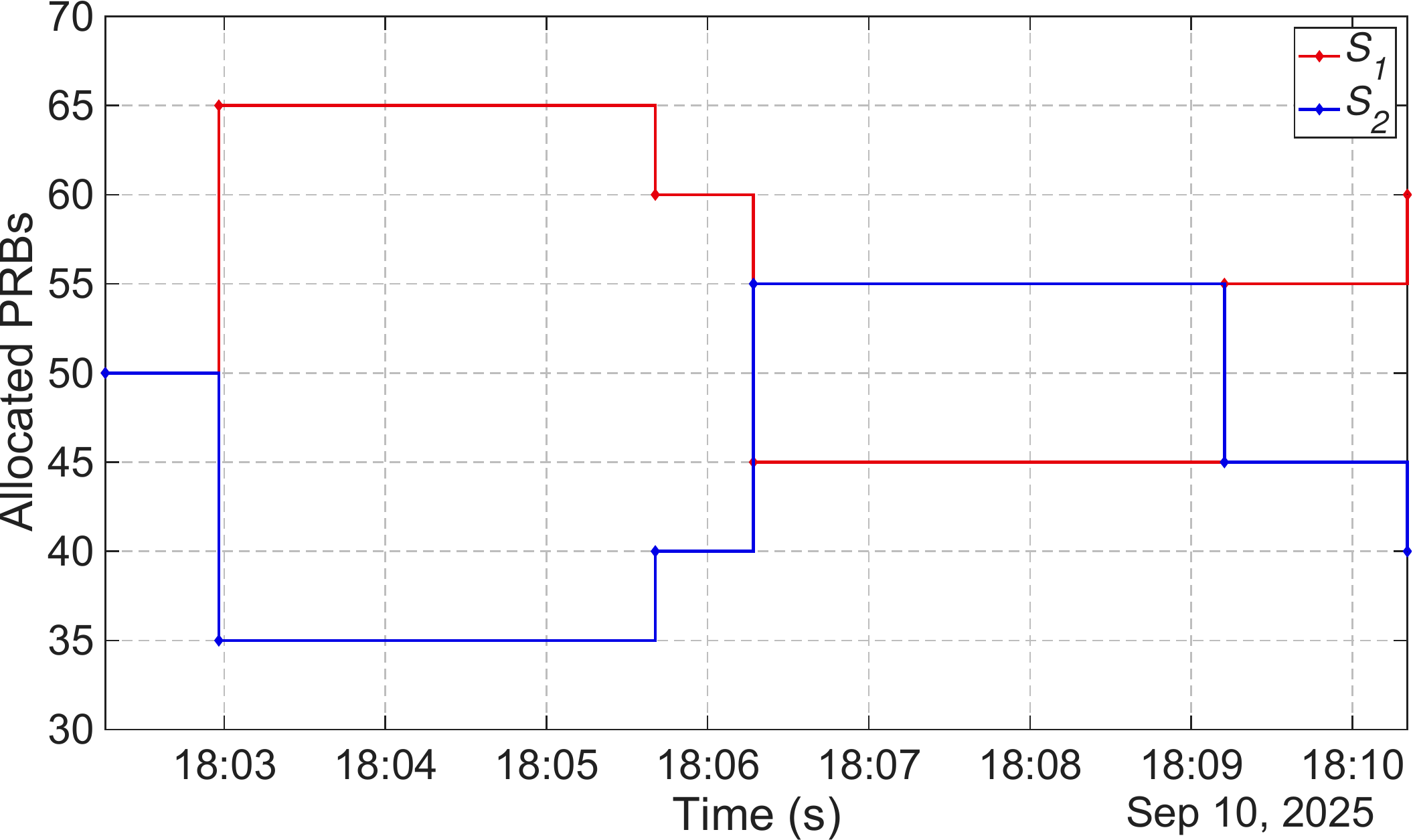}
        \end{subfigure}
        \begin{subfigure}{\columnwidth}
            \includegraphics[width=\textwidth, ]{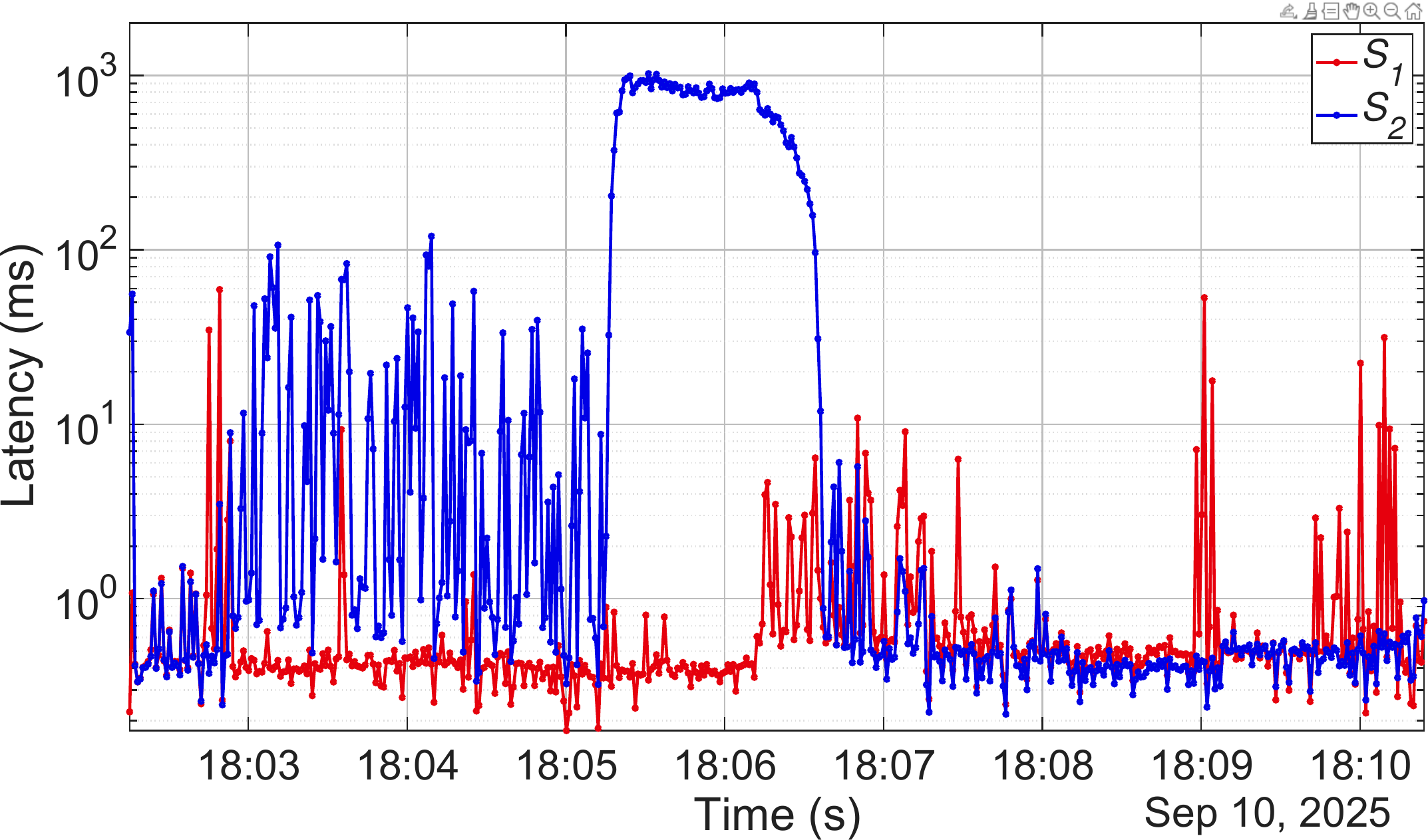}
        \end{subfigure}
        \caption{Illustration of dynamic radio resource allocation by the ReLLM under varying traffic conditions.}
        \label{latency_plot}
    \end{figure}
\subsection{Experimental Results}
We performed two distinct experimental cases that demonstrate the framework's capability to adaptively manage radio resources and maintain SLA compliance under varying network conditions.

    \subsubsection{Scenario 1}
    In this scenario, we illustrate a single continuous runtime instance of the system under dynamic traffic variations.
    Figure \ref{latency_plot} illustrates the adaptive behavior of the ReLLM in allocating radio resources, where traffic conditions dynamically vary over time. The initial traffic rates of both slices are set to approximately $80$ Mbps. When the traffic of $S_1$ is gradually increased to around $120$ Mbps, the available RBs become insufficient to meet the growing demand, which increases the latency of $S_1$. The resource allocation agent is triggered after each inference cycle and decides on a new allocation rate. It can be clearly seen that the new allocation reduces the latency below the threshold while avoiding significant impact on the performance of $S_2$. Subsequently, the traffic for $S_1$ is reduced, whereas the traffic for $S_2$ is gradually increased. At each step, $S_2$ initially experiences packet drops, but two reallocations by the resource allocation agent help maintain its drop ratio at zero. In the last phase, $S_2$ traffic is reduced while $S_1$ traffic increases again, then the agent intervenes once more to control latency. Overall, the result confirms that the ReLLM efficiently adapts to the dynamic traffic patterns and sustains system performance autonomously. 
    
    \subsubsection{Scenario 2} 
     In this case, the traffic rates are randomly selected from the list of $[80, 85, 90, ..., 125]$ Mbps, and the experiment is repeated $70$ times for statistical reliability.
    \begin{figure}[t]
        \centering
        \begin{minipage}{1\linewidth}
            \centering
            \subfloat[]{%
                \includegraphics[width=\linewidth]{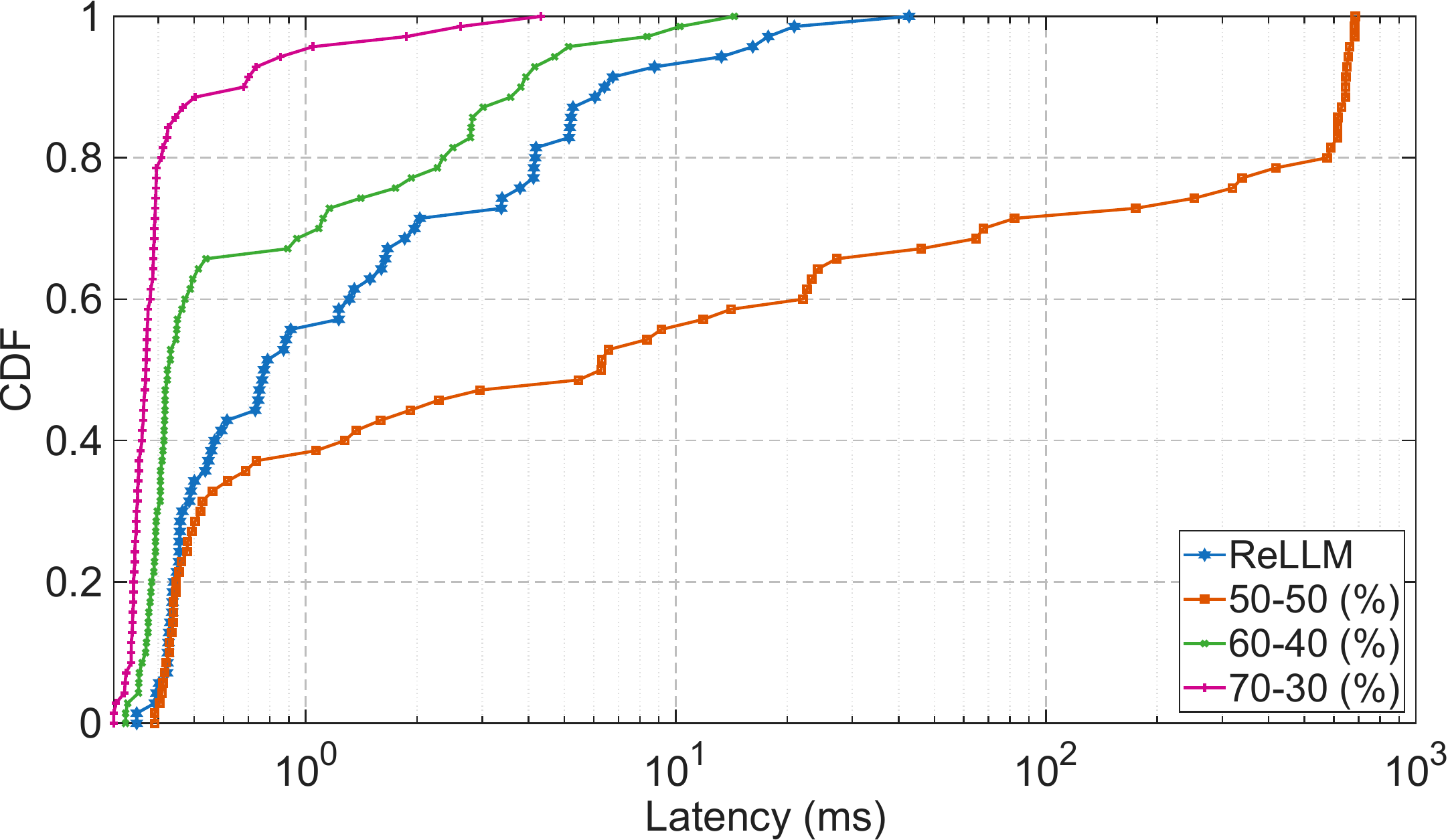}%
            }
        \end{minipage}
        \vspace{0.5em} 
    
        \begin{minipage}{1\linewidth}
            \centering
            \subfloat[]{%
                \includegraphics[width=\linewidth]{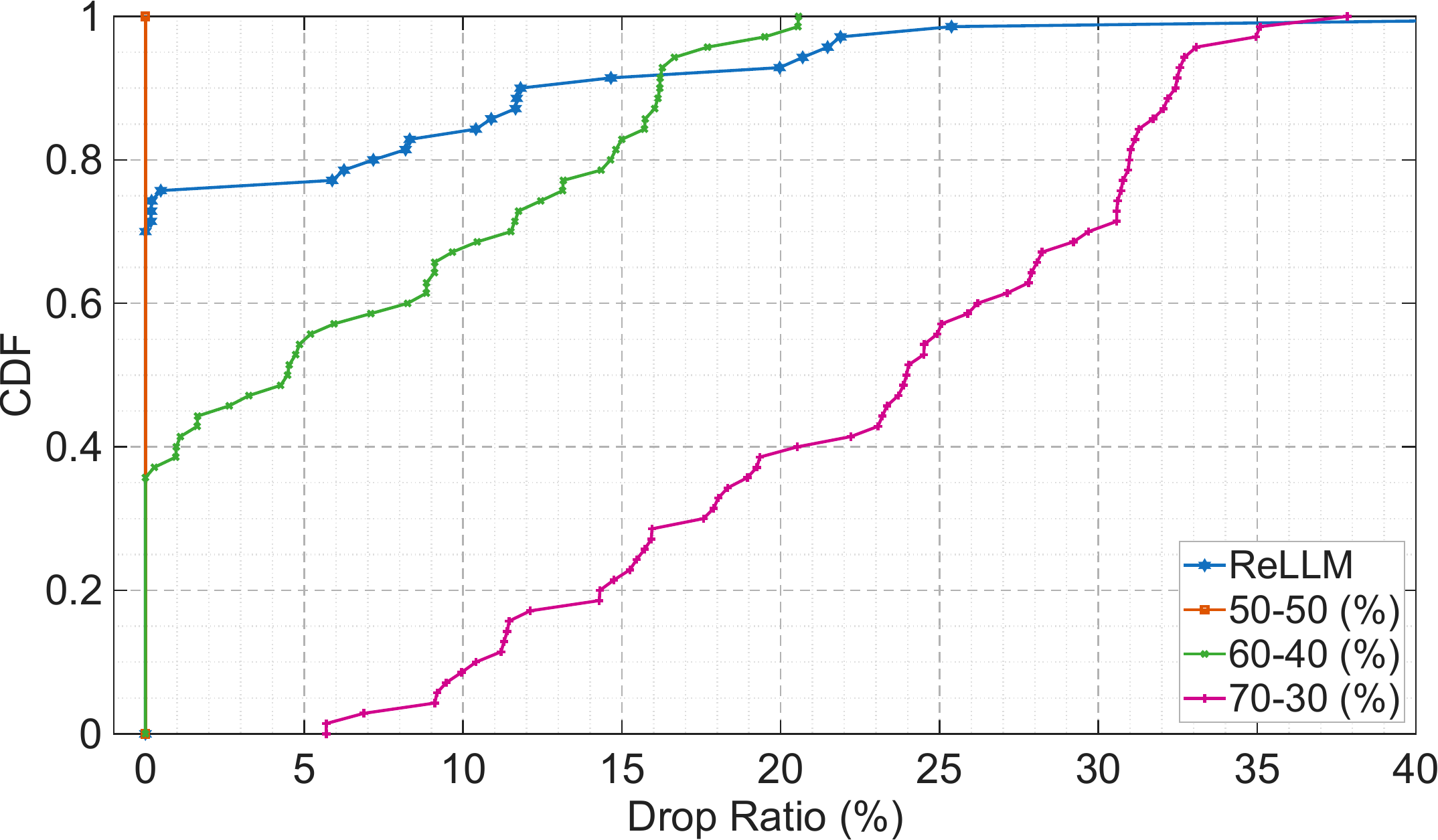}%
            }
        \end{minipage}
    
        \caption{CDF of (a) the latency and (b) the packet drop ratio for ReLLM-based and fixed resource allocation strategies.}
        \label{fig:cdfLatency}
    \end{figure}
  Fig. \ref{fig:cdfLatency} demonstrates that the LLM-based resource allocation technique provides a more balanced trade-off between high-priority and low-priority slices compared to fixed allocation strategies ($50-50$, $60-40$, $70-30$). While the $70-30$ configuration achieves the lowest latency for the high-priority slice, it does so at the expense of severe packet drops in the low-priority slice, reaching up to $50\%$. Conversely, the LLM approach significantly reduces this unfairness by maintaining a near-zero drop ratio for the low-priority slice while still ensuring acceptable latency performance for the high-priority slice. This highlights the ability of the LLM technique to adaptively allocate resources in a manner that preserves both reliability and fairness across heterogeneous service requirements in O-RAN systems.

\begin{figure}[t]
    \centering
    \begin{minipage}{1\linewidth}
        \centering
        \subfloat[]{%
            \includegraphics[width=\linewidth]{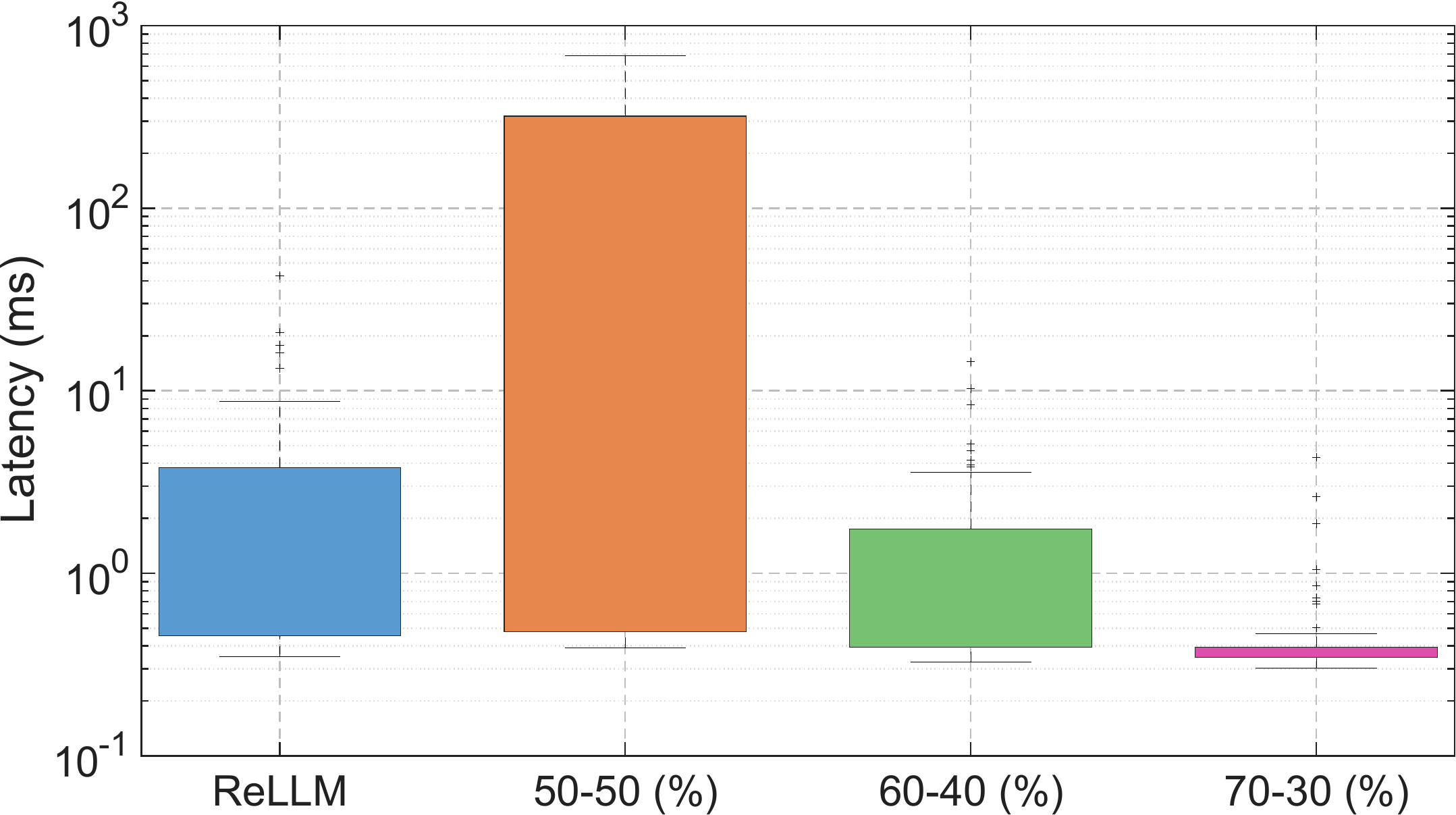}%
            \label{fig:latencybox}%
        }
    \end{minipage}
    \vspace{0.5em} 

    \begin{minipage}{1\linewidth}
        \centering
        \subfloat[]{%
            \includegraphics[width=\linewidth]{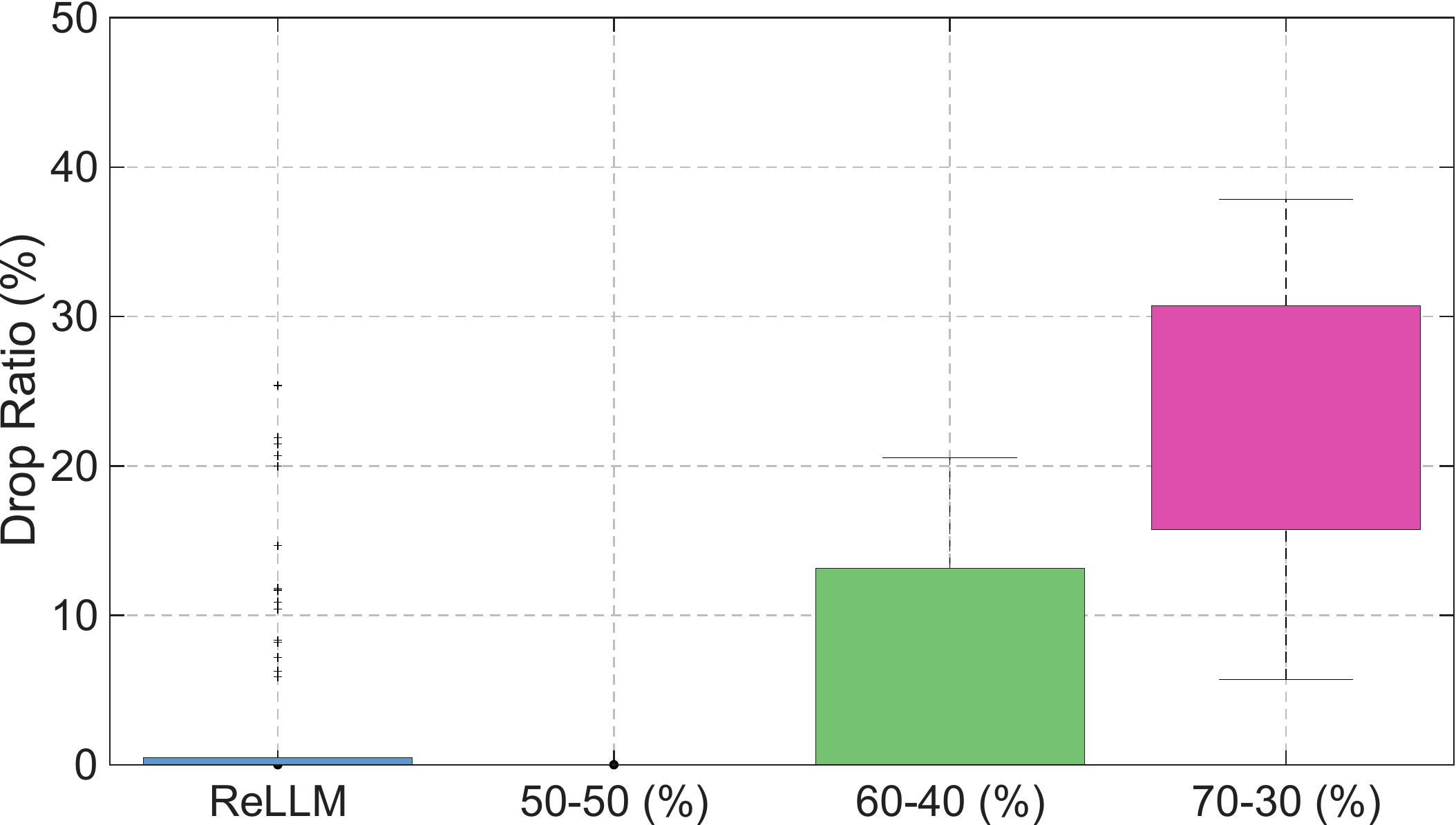}%
            \label{fig:dropratiobox}%
        }
    \end{minipage}

    \caption{Performance analysis of the proposed ReLLM-based resource allocation technique against fixed allocation strategies}
    \label{fig:latency_boxplot}
\end{figure}
Finally, Fig. \ref{fig:latency_boxplot} presents the results of Fig. \ref{fig:cdfLatency} in another format to show that the SLA of both slices are successfully satisfied. ReLLM effectively maintains the latency of $S_1$ under 10 ms while providing a drop ratio about $0.5\%$ for $S_2$. For instance, while the $70-30$ allocation provides the lowest latency for the high-priority slice, it causes a high packet drop rate of up to $50\%$ for the low-priority slice. 
On the other hand, the LLM-based approach maintains a near-zero drop ratio for the low-priority slice, while still keeping the latency within acceptable limits for the high-priority slice. The results confirm that the LLM method provides adaptive and efficient resource allocation, ensuring both fairness and service reliability under heterogeneous requirements.

\subsection{Complexity Analysis}
\begin{figure}
        \centering
        \includegraphics[width=\linewidth]{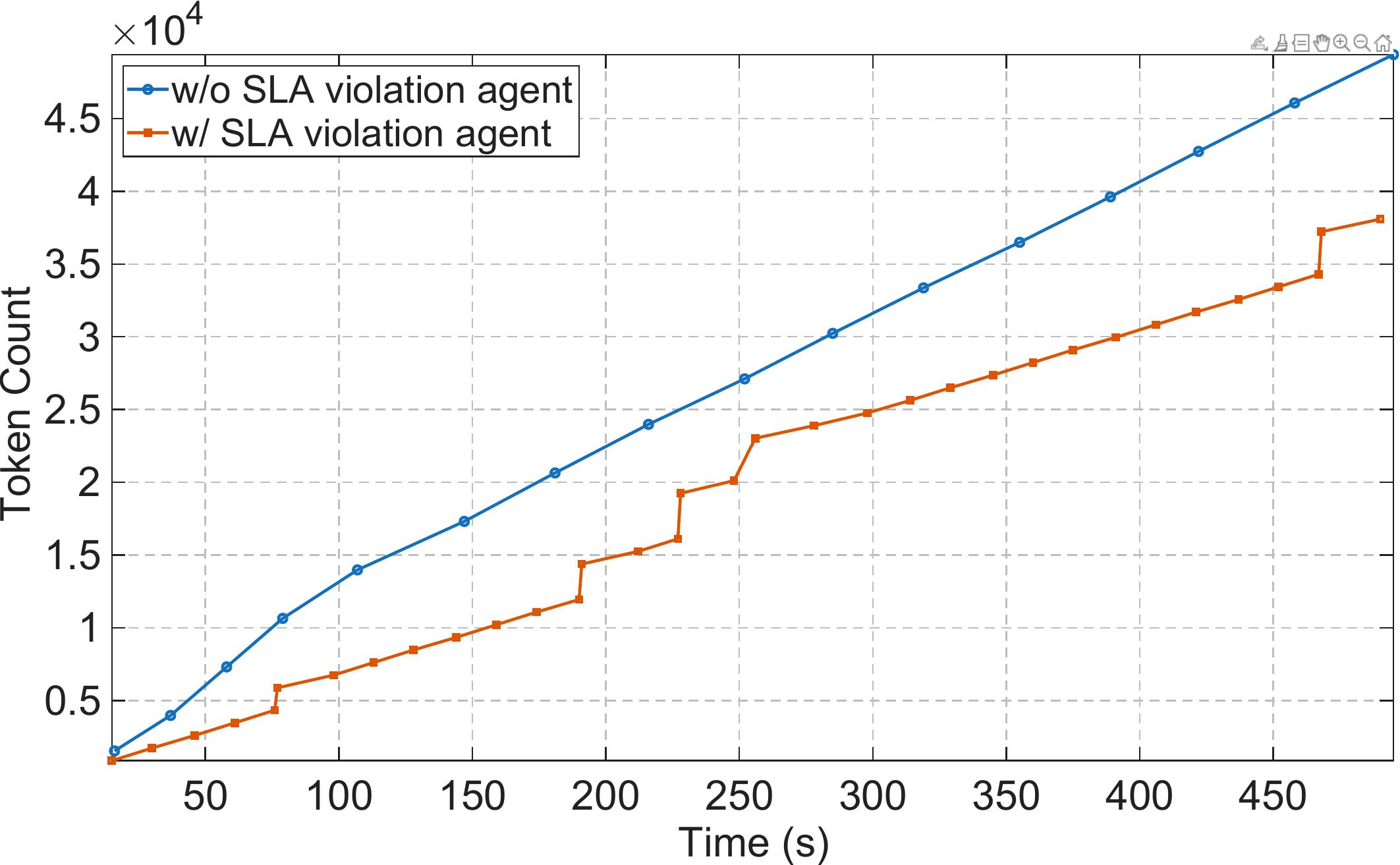}
        \caption{Comparison of ReLLM token usage with and without the SLA violation agent.}
        \label{fig:token}
    \end{figure}

In large-scale O-RAN systems, computational complexity and energy efficiency are critical factors for the real-time applicability of AI-driven resource management frameworks. In the proposed ReLLM framework, the RAG mechanism shortens the effective context length by retrieving only the most relevant historical data for each decision step. Also, the introduction of an SLA violation detection agent ensures that full-scale LLM inference is triggered only when a potential SLA violation is detected, thereby avoiding unnecessary attention computations.
Figure \ref{fig:token} compares the token consumption of two ReLLM-based resource allocation approaches under Scenario 1, where one approach uses an SLA violation detection agent and the other makes decisions at every step without this agent. The result shows that the SLA violation detection agent significantly reduces the token consumption over time by reducing the frequency of resource allocation agent executions,
which are more costly due to its longer meta-prompt. This behavior demonstrates that the proposed dual-agent ReLLM architecture not only maintains decision accuracy but also achieves higher computational efficiency and energy awareness.
    
\section{Conclusion}
In this paper, we present a RAG-empowered LLM framework for SLA-aware dynamic radio resource management among network slices with diverse QoS requirements in O-RAN inspired network. The framework employs two specialized agents, namely an SLA violation detection agent and a resource allocation agent. The SLA violation detection agent continuously observes and predicts SLA risks, while the resource allocation agent dynamically distributes RBs to maintain SLA compliance and optimize network performance under varying traffic conditions. The framework is implemented and validated on an end-to-end OAI-based O-RAN testbed to assess system accuracy and adaptability. The results show that the ReLLM-based approach consistently maintained the high-priority slice below the 10 ms latency threshold while maximizing the performance of the secondary slice. Also, the system successfully adapted RBs allocations in response to traffic fluctuations. Overall, the results indicate that the ReLLM-driven approach provides a promising alternative to traditional AI-based approaches. Through this work, it is demonstrated that the utilization of LLMs for radio resource management in O-RAN compliant network is both feasible and effective. Future researches focus on extending the framework with multi-agent network configuration, designing lightweight and energy-aware LLM models tailored for edge-deployed RIC environments, exploring the use of LLMs for proactive fault detection, anomaly reasoning, and security policy generation in O-RAN systems.

\section*{Acknowledgment}
The authors would like to thank Prof. Dr. Halim Yanikomeroglu for his valuable comments and insightful suggestions. This paper has received funding from the MOSAIC project. MOSAIC has been accepted for funding within the CHIPS Joint Undertaking, a public-private partnership in collaboration with the HORIZON Framework Programme and the national Authorities under grant agreement number 101194414.

\bibliographystyle{IEEEtran}
\bibliography{main.bib}
\end{document}